\documentstyle[multicol,aps,psfig]{revtex}
\begin{document}

\title{Multiple Parton Scattering in Nuclei: Beyond Helicity Amplitude
Approximation}
\author{Ben-Wei Zhang$^{a}$ and Xin-Nian Wang$^{b,c}$}
\address{$^a$ Institute of Particle Physics, Huazhong Normal University,
         Wuhan 430079, China}
\address{$^b$Nuclear Science Division, MS 70R0319,
Lawrence Berkeley National Laboratory, Berkeley, CA 94720 USA}
\address{$^c$Department of Physics, Shandong University,
         Jinan 250100, China}


\maketitle


\begin{abstract}
\baselineskip=12pt Multiple parton scattering and induced parton
energy loss in deeply inelastic scattering (DIS) off heavy nuclei
is studied within the framework of generalized factorization in
perturbative QCD with a complete calculation beyond the helicity
amplitude (or soft bremsstrahlung) approximation. Such a
calculation gives rise to new corrections to the modified quark
fragmentation functions. The effective parton energy loss is found
to be reduced by a factor of 5/6 from the result of helicity
amplitude approximation.
\end{abstract}


\baselineskip=16pt

\section{Introduction}

Suppression of jet production or jet quenching in high-energy
nuclear collisions has been proposed as a good probe of the hot
and dense medium \cite{GP,WG} that is produced during the violent
collisions. The quenching of an energetic parton is caused by
multiple scattering and induced parton energy loss during its
propagation through the hot QCD medium. It
suppresses the final leading hadron distribution giving rise to
modified fragmentation functions and the final hadron spectra
\cite{WH,cw}. Recent theoretical estimates
\cite{GW1,BDMPS,Zh,GLV,Wie} all show that the effective parton
energy loss is proportional to the gluon density of the medium.
Therefore measurements of the parton energy loss will enable one to
extract the initial gluon density of the produced hot medium.
Strong suppression of high transverse momentum hadron spectra is
indeed observed by experiments \cite{Phenix,Star} at the
Relativistic Heavy-Ion Collider (RHIC) at the Brookhaven National
Laboratory (BNL), indicating large parton energy loss in a medium
with large initial gluon density. However, one cannot
unambiguously extract the initial gluon density from the
experiments of heavy-ion collisions alone because of the
theoretical uncertainty in relating the parton energy loss to the
initial gluon density. For this purpose, one has to rely on other
complimentary experimental measurements such as parton energy loss
in deeply inelastic scattering (DIS) of nuclear targets. One can
then at least extract the initial gluon density in heavy-ion
collisions relative to that in a cold nucleus \cite{EW1}.

Modified quark fragmentation function inside a nucleus in DIS and
the effective parton energy loss has been derived recently by Guo and
Wang \cite{GW}. Generalized factorization of twist-four
processes \cite{LQS} was applied to the inclusive process of jet
fragmentation in DIS in order to derive the modified fragmentation
functions. Taking into account of gluon bremsstrahlung induced by
multiple parton scattering and the Landau-Pomeranchuck-Migdal (LPM)
interference effect, one finds that the leading twist-four
contributions to the modified fragmentation function and the
effective parton energy loss depend quadratically on the nuclear
size $R_A$. They also depend linearly on the effective gluon
distribution in nuclei. One can also extend the study to parton
propagation inside a hot QCD medium reproducing earlier results \cite{EW1}.
This allows one to relate parton energy loss in both hot and cold
nuclear medium.

There are all together 23 cut-diagrams that contribute to the
leading twist-four corrections to the quark fragmentation
function in $eA$ DIS. For simplification of the calculation,
the helicity amplitude approximation was used in Ref.~\cite{GW}
in the limit of soft gluon radiation $z_g=1-z \rightarrow 0 $
where $z_g$ is the momentum fraction carried by the radiated
gluon and $z$ the fraction carried by the leading quark.
Such an approximation enables one to simplify the calculation of
the radiation amplitudes.
The final results are obtained by squaring the sum of all
possible amplitudes, giving rise not only to the contributions
of double scattering but also various interferences.
In this approximation, the amplitudes of initial and final state
radiation are the same except the opposite signs and different
color matrices. Because of the different color matrices in the
initial and final state radiation, there is no complete
cancellation of the radiation amplitudes. In addition, there
is also gluon radiation from the exchanged gluon via triple-gluon coupling.
These non-Abelian features of QCD radiation lead to a finite
gluon spectra even in the helicity amplitude approximation.
However, under the same approximation, the photon spectra
from QED bremsstrahlung would be zero because of almost complete
cancellation between initial and final state radiation.
One therefore has to go beyond the helicity approximation.
In this paper, we will study the correction to the gluon radiation
spectra when we go beyond the helicity amplitude approximation
and its effect in the modified quark fragmentation function.
We will also compute the
effective quark energy loss and compare to the result in the
helicity amplitude approximation.


\section{Generalized Factorization}
In order to study the quark fragmentation in $eA$ DIS,
we consider the following semi-inclusive processes,
$e(L_1) + A(p) \longrightarrow e(L_2) + h (\ell_h) +X$,
where $L_1$ and $L_2$ are the four momenta of the incoming and the
outgoing leptons, and $\ell_h$ is the observed hadron momentum.
The differential
cross section for the semi-inclusive process can be expressed as
\begin{equation}
E_{L_2}E_{\ell_h}\frac{d\sigma_{\rm DIS}^h}{d^3L_2d^3\ell_h}
=\frac{\alpha^2_{\rm EM}}{2\pi s}\frac{1}{Q^4} L_{\mu\nu}
E_{\ell_h}\frac{dW^{\mu\nu}}{d^3\ell_h} \; ,
\label{sigma}
\end{equation}
where $p = [p^+,0,{\bf 0}_\perp] \label{eq:frame}$
is the momentum per nucleon in the nucleus,
$q =L_2-L_1 = [-Q^2/2q^-, q^-, {\bf 0}_\perp]$ the momentum transfer,
$s=(p+L_1)^2$ and $\alpha_{\rm EM}$ is the electromagnetic (EM)
coupling constant. The leptonic tensor is given by
$L_{\mu\nu}=1/2\, {\rm Tr}(\gamma \cdot L_1 \gamma_{\mu}
\gamma \cdot L_2 \gamma_{\nu})$
while the semi-inclusive hadronic tensor is defined as,
\begin{eqnarray}
E_{\ell_h}\frac{dW_{\mu\nu}}{d^3\ell_h}&=&
\frac{1}{2}\sum_X \langle A|J_\mu(0)|X,h\rangle
\langle X,h| J_\nu(0)|A\rangle \nonumber \\
&\times &2\pi \delta^4(q+p-p_X-\ell_h)
\end{eqnarray}
where $\sum_X$ runs over all possible final states and
$J_\mu=\sum_q e_q \bar{\psi}_q \gamma_\mu\psi_q$ is the
hadronic EM current.

In the parton model with collinear factorization approximation,
the leading-twist contribution to the semi-inclusive cross section
can be factorized into a product of parton distributions,
parton fragmentation functions and the partonic cross section.
Including all leading log radiative corrections, the lowest order
contribution (${\cal O}(\alpha_s^0)$) from a single
hard $\gamma^*+ q$ scattering can be written as
\begin{eqnarray}
& &\frac{dW^S_{\mu\nu}}{dz_h}
= \sum_q e_q^2 \int dx f_q^A(x,\mu_I^2) H^{(0)}_{\mu\nu}(x,p,q)
D_{q\rightarrow h}(z_h,\mu^2)\, ; \label{Dq} \\
& &H^{(0)}_{\mu\nu}(x,p,q) = \frac{1}{2}\,
{\rm Tr}(\gamma \cdot p \gamma_{\mu} \gamma \cdot(q+xp) \gamma_{\nu})
\, \frac{2\pi}{2p\cdot q} \delta(x-x_B) \, , \label{H0}
\end{eqnarray}
where the momentum fraction carried by the hadron is defined as
$z_h=\ell_h^-/q^-$ and $x_B=Q^2/2p^+q^-$ is the Bjorken variable.
$\mu_I^2$ and $\mu^2$ are the factorization scales for the initial
quark distributions $f_q^A(x,\mu_I^2)$ in a nucleus and the fragmentation
functions $D_{q\rightarrow h}(z_h,\mu^2)$, respectively.
The renormalized quark fragmentation function
$D_{q\rightarrow h}(z_h,\mu^2)$ satisfies the
Dokshitzer-Gribov-Lipatov-Altarelli-Parisi (DGLAP) QCD evolution
equations \cite{AP}.

In a nuclear medium, the propagating quark in DIS will experience additional
scatterings with other partons from the nucleus. The rescatterings may
induce additional gluon radiation and cause the leading quark to lose
energy. Such induced gluon radiations will effectively give rise to
additional terms in the evolution equation leading to the modification of the
fragmentation functions in a medium. These are so-called higher-twist
corrections since they involve higher-twist parton matrix elements and
are power-suppressed. We will consider those contributions that
involve two-parton correlations from two different nucleons inside
the nucleus. They are proportional to the size of the nucleus \cite{OW} 
and thus are
enhanced by a nuclear factor $A^{1/3}$ as compared to two-parton correlations
in a nucleon. Like in previous studies \cite{GW}, we will neglect
those contributions that are not enhanced by the nuclear medium.

\begin{figure}
\centerline{\psfig{file=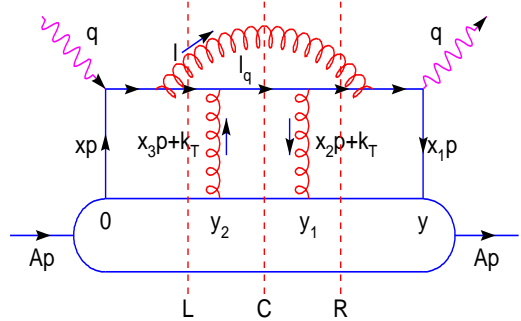,width=3in,height=2.0in}}
\caption{A typical diagram for quark-gluon re-scattering processes with three
possible cuts, central(C), left(L) and right(R).}
\label{fig1}
\end{figure}

We will employ the generalized factorization of multiple scattering
processes \cite{LQS}. In this approximation, the double scattering
contribution to radiative correction from processes like the one
illustrated in Fig.~\ref{fig1} can be written in the following form,
\begin{eqnarray}
\frac{dW_{\mu\nu}^D}{dz_h}
&=& \sum_q \,\int_{z_h}^1\frac{dz}{z}D_{q\rightarrow h}(z_h/z)
   \int \frac{dy^{-}}{2\pi}\, dy_1^-dy_2^-
\frac{1}{2}\,
     \langle A | \bar{\psi}_q(0)\,
           \gamma^+\, F_{\sigma}^{\ +}(y_{2}^{-})\,
 F^{+\sigma}(y_1^{-})\,\psi_q(y^{-})
     | A\rangle \nonumber \\
&\times &
    \left(-\frac{1}{2}g^{\alpha\beta}\right)
\left[\, \frac{1}{2}\, \frac{\partial^{2}}
                   {\partial k_{T}^{\alpha} \partial k_{T}^{\beta}}\,
    \overline{H}^D_{\mu\nu}(y^-,y_1^-,y^-_2,k_T,p,q,z)\, \right]_{k_T=0}\, ,
\label{factorize}
\end{eqnarray}
after collinear expansion of the hard partonic cross section with
respect to the transverse momentum of the initial partons,
where $\overline{H}^D_{\mu\nu}(y^-,y_1^-,y^-_2,k_T,p,q,z)$ is the
Fourier transform of the partonic hard part
$\widetilde{H}_{\mu\nu}(x,x_1,x_2,k_T,p,q,z)$ in momentum space,
\begin{eqnarray}
\overline{H}^D_{\mu\nu}(y^-,y_1^-,y^-_2,k_T,p,q,z)
&=& \int dx\, \frac{dx_1}{2\pi}\, \frac{dx_2}{2\pi}\,
       e^{ix_1p^+y^- + ix_2p^+y_1^- + i(x-x_1-x_2)p^+y_2^-}\nonumber \\
&\ & \times \widetilde{H}^D_{\mu\nu}(x,x_1,x_2,k_T,p,q,z)\ ,
\label{factorize-1}
\end{eqnarray}
and $k_T$ is the relative transverse momentum carried by the
second parton in the double scattering. This is the leading
term in the collinear expansion that contributes to the double
scattering process. The first term in the collinear expansion gives
the eikonal contribution to the leading-twist results, making the
matrix element in the single scattering process gauge invariant, 
while the second (or linear) term vanishes for unpolarized 
initial and final states after integration over $k_T$.

The hard part of the partonic scattering for each diagram,
$\widetilde{H}_{\mu\nu}(x,x_1,x_2,k_T,p,q,z)$, always contains two
$\delta$-functions from the on-shell conditions of the two
cut-propagators. These $\delta$-functions, together with the
contour integrations which contain different sets of poles in the
un-cut propagators, will determine the values of the momentum
fractions $x, x_1,$ and $x_2$ \cite{GW}. The phase factors in
$\overline{H}^D_{\mu\nu}(y^-,y_1^-,y^-_2,k_T,p,q,z)$
[Eq.~(\ref{factorize-1})] can then be factored out, which will be
combined with the partonic fields in Eq.~(\ref{factorize}) to form
twist-four partonic matrix elements or two-parton correlations.
The double scattering corrections in Eq.~(\ref{factorize}) can
then be factorized into the product of fragmentation functions,
twist-four partonic matrix elements and the partonic hard
scattering cross section.

\section{Beyond Helicity Amplitude Approximation}

To simplify the calculation of various cut-diagrams of double
scattering and illustrate the underlying physical processes,
helicity amplitude approximation was used in Ref.~\cite{GW}. In
this approximation, one neglects the transverse recoil induced by
the scattering and consider only the part of the amplitudes in
which quarks' helicity is unchanged in the scattering. The final
results will agree with the complete calculation in the limit of
soft radiation.

Take photon bremsstrahlung for example. A complete calculation of
photon radiation induced by a single scattering with transverse momentum
transfer $k_T$ gives a spectra
\begin{equation}
\frac{dN}{d\ell_T^2 dz}=\frac{\alpha}{2\pi}
\left[\frac{\vec{\ell}_T}{\ell_T^2}
-\frac{\vec{\ell}_T+(1-z)\vec{k}_T}{(\vec{\ell}_T+(1-z)\vec{k}_T)^2}\right]^2
\frac{1+z^2}{1-z} \;\; .
\end{equation}
Here we denote the momentum of the photon (or gluon in QCD) to be
$\ell$ which carries $1-z$ momentum fraction of the struck quark.
Under helicity amplitude approximation, the splitting function
will become $2/(1-z)$ and furthermore the term $(1-z)k_T$ in the
final state radiation amplitude will be neglected. The
interference between initial and final state radiation will
effectively reduce the photon radiation spectrum to zero,
apparently not a precise approximation. In QCD, the corresponding
gluon radiation has similar amplitudes, except additional color
matrices. Because the gluon exchange in the scattering transfers
color, the emitted gluon in the initial and final state radiation
can carry different colors. In this case, there is no complete
destructive interference between initial and final state radiation
as in QED. The helicity amplitude approximation is, therefore, a
better approximation in QCD than in QED. However, it still
neglects the corrections which contribute the most in photon
radiation in QED. This finite correction is what we will study in
this paper.

We first consider the contribution from Fig.~\ref{fig1} in detail and
will list the results of other diagrams afterwards. Using the
conventional Feynman rule, one can write down the hard partonic
part of the central cut-diagram of Fig.~\ref{fig1} \cite{GW},
\begin{eqnarray}
\overline{H}^D_{C\,\mu\nu}(y^-,y_1^-,y_2^-,k_T,p,q,z)&=&
\int dx\frac{dx_1}{2\pi}\frac{dx_2}{2\pi}
e^{ix_1p^+y^- + ix_2p^+y_1^- + i(x-x_1-x_2)p^+y_2^-}
\int \frac{d^4\ell}{(2\pi)^4} \nonumber \\
&\times&\frac{1}{2}{\rm Tr}\left[p\cdot\gamma\gamma_\mu p^\sigma p^\rho
\widehat{H}_{\sigma\rho}\gamma_\nu \right]
2\pi\delta_+(\ell^2)\,
\delta(1-z-\frac{\ell^-}{q^-}) \; . \label{eq:fig1-1}
\end{eqnarray}

\begin{eqnarray}
\widehat{H}_{\sigma\rho} &=&
\frac{C_F}{2N_c}g^4\frac{\gamma\cdot(q+x_1 p)}{(q+x_1p)^2-i\epsilon}
\,\gamma_\alpha\,\frac{\gamma\cdot(q + x_1 p
-\ell)}{(q+x_1p-\ell)^2-i\epsilon}
\,\gamma_\sigma \gamma\cdot\ell_q\,\gamma_\rho
\nonumber \\
&\times &\varepsilon^{\alpha\beta}(\ell)\frac{\gamma\cdot(q+xp -\ell)}
{(q+xp-\ell)^2+i\epsilon} \,\gamma_\beta\,
\frac{\gamma\cdot(q + xp)}{(q+xp)^2+i\epsilon}
\,\,2\pi \delta_+(\ell_q^2)\, , \label{eq:fig1-2}
\end{eqnarray}
where $\varepsilon^{\alpha\beta}(\ell)$ is the polarization tensor of a
gluon propagator in an axial gauge, $n\cdot A=0$ with
$n=[1,0^-,\vec{0}_\perp]$, and
$\ell$, $\ell_q=q+(x_1+x_2)p+k_T-\ell$
are the 4-momenta carried by the gluon and the final quark, respectively.
$z=\ell_q^-/q^-$ is the fraction of longitudinal momentum
(the large minus component) carried by the final quark.

To simplify the calculation, we also apply the collinear
approximation to complete the trace of the product of
$\gamma$-matrices,
\begin{equation}
p^\sigma\widehat{H}_{\sigma\rho}p^\rho
\approx \gamma\cdot\ell_q \,\frac{1}{4\ell_q^-}
{\rm Tr} \left[\gamma^- p^\sigma\widehat{H}_{\sigma\rho}p^\rho\right] \; .
\label{coll}
\end{equation}
After carrying out momentum integration in $x$, $x_1$, $x_2$
and $\ell^{\pm}$ with the help of contour integration
and $\delta$-functions, the partonic hard part
can be factorized into the production of $\gamma$-quark scattering
matrix $H^{(0)}_{\mu\nu}(x,p,q)$ [Eq.~(\ref{H0})]
and the quark-gluon rescattering part $\overline{H}^D$,
\begin{equation}
\overline{H}^D_{\mu\nu}(y^-,y_1^-,y_2^-,k_T,p,q,z) =
\int dx H^{(0)}_{\mu\nu}(x,p,q)\
\overline{H}^D(y^-,y_1^-,y_2^-,k_T,x,p,q,z)\, . \label{eq:hc0}
\end{equation}
Contributions from all the diagrams have this factorized from.
Therefore, we will only list the rescattering part
$\overline{H}^D$ for different diagrams in the following. For the
central-cut diagram in Fig.~\ref{fig1} it reads \cite{GW},
\begin{eqnarray}
\overline{H}^D_{C(Fig.\ref{fig1}) }(y^-,y_1^-,y_2^-,k_T,x,p,q,z)&=&
\int \frac{d\ell_T^2}{\ell_T^2}\, \frac{\alpha_s}{2\pi}\,
 C_F\frac{1+z^2}{1-z} \nonumber \\
&\times&\frac{2\pi\alpha_s}{N_c}
\overline{I}_{C(Fig.\ref{fig1}) }(y^-,y_1^-,y_2^-,\ell_T,k_T,x,p,q,z)
 \, , \label{eq:hc1}
\end{eqnarray}
\begin{eqnarray}
\overline{I}_{C(Fig.\ref{fig1}) }(y^-,y_1^-,y_2^-,\ell_T,k_T,x,p,q,z)
&=&e^{i(x+x_L)p^+y^- + ix_Dp^+(y_1^- - y_2^-)}
\theta(-y_2^-)\theta(y^- - y_1^-) \nonumber \\
&\times &(1-e^{-ix_Lp^+y_2^-})(1-e^{-ix_Lp^+(y^- - y_1^-)}) \; .
\label{eq:Ic1}
\end{eqnarray}
Here, the fractional momentum is defined as
\begin{eqnarray}
  x_L&=&\frac{\ell_T^2}{2p^+q^-z(1-z)} \,\, ,\,\,
  x_D=\frac{k_T^2-2\vec{k}_T\cdot \vec{\ell}_T}{2p^+q^-z} \, ,
\label{xld}
\end{eqnarray}
and $x=x_B=Q^2/2p^+q^-$ is the Bjorken variable.

The above contribution resembles the cross section of dipole scattering
and contains essentially four terms. The first diagonal term
corresponds to the so-called hard-soft process where the
gluon radiation is induced by the hard scattering between the virtual photon
and an initial quark with momentum fraction $x$. The quark is
knocked off-shell by the virtual photon and becomes on-shell again after
radiating a gluon. Afterwards the on-shell 
quark (or the radiated gluon) will have a
secondary scattering with another soft gluon from the nucleus.
The second diagonal term is due to the so-called double hard process
where the quark is on-shell after the first hard scattering with the
virtual photon. The gluon radiation is then induced by the scattering of
the quark with another gluon that carries finite momentum fraction $x_L+x_D$.
The other two off-diagonal terms are interferences between hard-soft
and double hard processes. In the limit of collinear
radiation ($x_L\rightarrow 0$) or when the formation time of the
gluon radiation, $\tau_f\equiv 1/x_Lp^+$, is much larger
than the nuclear size, the two processes have destructive interference,
leading to the LPM interference effect.

\begin{figure}
\centerline{\psfig{file=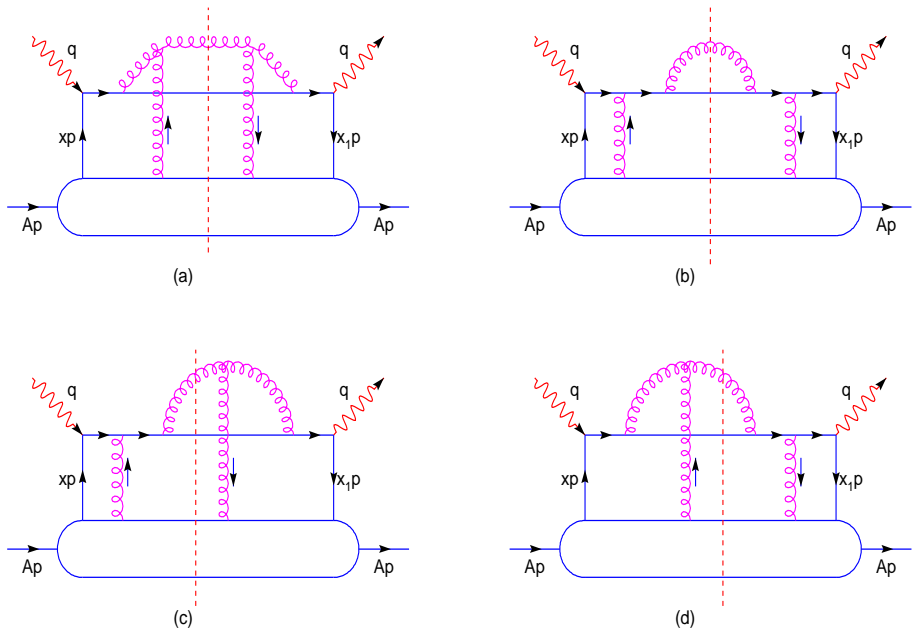,width=5.in,height=3.5in}}
\caption{Four central-cut Diagrams that contribute to the final results.}
\label{fig2}
\end{figure}

One can similarly obtain the  rescattering part $\overline{H}^D$
of other central-cut diagrams (a-d) in Fig.~\ref{fig2}:


\begin{eqnarray}
\overline{H}^D_{C(a)}(y^-,y_1^-,y_2^-,k_T,x,p,q,z)&=&
\int \frac{d\ell_T^2}{(\vec{\ell_T}-\vec{k_T})^2}\, \frac{\alpha_s}{2\pi}\,
 C_A\frac{1+z^2}{1-z} \nonumber \\
&\times&\frac{2\pi\alpha_s}{N_c}
\overline{I}_{C(a)}(y^-,y_1^-,y_2^-,\ell_T,k_T,x,p,q,z)
 \, , \nonumber \\
\overline{I}_{C(a)}(y^-,y_1^-,y_2^-,\ell_T,k_T,x,p,q,z)&=&
e^{i(x+x_L)p^+y^-+ix_Dp^+(y_1^--y_2^-)}
\theta(-y_2^-)\theta(y^- - y_1^-) \nonumber \\
&\times&[e^{ix_Dp^+y_2^-/(1-z)}-e^{-ix_Lp^+y_2^-}] \nonumber \\
&\times&[e^{ix_Dp^+(y^- - y_1^-)/(1-z)}-e^{-ix_Lp^+(y^- - y_1^-)}] \, ,
\label{eq:hc(a)}
\end{eqnarray}

\begin{eqnarray}
\overline{H}^D_{C(b)}(y^-,y_1^-,y_2^-,k_T,x,p,q,z)&=&
\int \frac{d\ell_T^2}{(\vec{\ell_T}-(1-z)\vec{k_T})^2}\, \frac{\alpha_s}{2\pi}\,
 C_F\frac{1+z^2}{1-z} \nonumber \\
&\times&\frac{2\pi\alpha_s}{N_c}
\overline{I}_{C(b)}(y^-,y_1^-,y_2^-,\ell_T,k_T,x,p,q,z)
 \, , \nonumber \\
\overline{I}_{C(b)}(y^-,y_1^-,y_2^-,\ell_T,k_T,x,p,q,z)&=&
e^{i(x+x_L)p^+y^-+ix_Dp^+(y_1^--y_2^-)}
\theta(-y_2^-)\theta(y^- - y_1^-) \nonumber \\
&\times& e^{-ix_Lp^+(y^- - y_1^-)}e^{-ix_Lp^+y_2^-} \, ,
\label{eq:hc(b)}
\end{eqnarray}

\begin{eqnarray}
\overline{H}^D_{C(c)}(y^-,y_1^-,y_2^-,k_T,x,p,q,z)&=&
\int d\ell_T^2\frac{ (\vec{\ell_T}-\vec{k_T})\cdot
(\vec{\ell_T}-(1-z)\vec{k_T}) }
{(\vec{\ell_T}-\vec{k_T})^2 (\vec{\ell_T}-(1-z)\vec{k_T})^2}\,
\frac{\alpha_s}{2\pi}\,
\frac{C_A}{2}\frac{1+z^2}{1-z} \nonumber \\
&\times&\frac{2\pi\alpha_s}{N_c}
\overline{I}_{C(c)}(y^-,y_1^-,y_2^-,\ell_T,k_T,x,p,q,z)
 \, , \nonumber \\
\overline{I}_{C(c)}(y^-,y_1^-,y_2^-,\ell_T,k_T,x,p,q,z)&=&
e^{i(x+x_L)p^+y^-+ix_Dp^+(y_1^--y_2^-)} \,
\theta(-y_2^-)\theta(y^- - y_1^-) \nonumber \\
&\times&e^{-ix_Lp^+y_2^-} \nonumber \\
&\times&[e^{ix_Dp^+(y^- - y_1^-)/(1-z)}-e^{-ix_Lp^+(y^- - y_1^-)}] \, ,
\label{eq:hc(c)}
\end{eqnarray}

\begin{eqnarray}
\overline{H}^D_{C(d)}(y^-,y_1^-,y_2^-,k_T,x,p,q,z)&=&
\int d\ell_T^2\frac{ (\vec{\ell_T}-\vec{k_T})\cdot
(\vec{\ell_T}-(1-z)\vec{k_T}) }
{(\vec{\ell_T}-\vec{k_T})^2 (\vec{\ell_T}-(1-z)\vec{k_T})^2}\,
\frac{\alpha_s}{2\pi}\,
\frac{C_A}{2}\frac{1+z^2}{1-z} \nonumber \\
&\times&\frac{2\pi\alpha_s}{N_c}
\overline{I}_{C(d)}(y^-,y_1^-,y_2^-,\ell_T,k_T,x,p,q,z)
 \, , \nonumber \\
\overline{I}_{C(d)}(y^-,y_1^-,y_2^-,\ell_T,k_T,x,p,q,z)&=&
e^{i(x+x_L)p^+y^-+ix_Dp^+(y_1^--y_2^-)} \,
\theta(-y_2^-)\theta(y^- - y_1^-) \nonumber \\
&\times&e^{-ix_Lp^+(y^- - y_1^-)} \nonumber \\
&\times&[e^{ix_Dp^+y_2^-/(1-z)}-e^{-ix_Lp^+y_2^-}] \, .
\label{eq:hc(d)}
\end{eqnarray}

To complete the calculation we also have to consider
the asymmetrical-cut diagrams(left cut and right cut) that represent
interferences between single and triple scatterings.
They can be obtained with similar procedures. We list
the rescattering part $\overline{H}^D$ of all those
asymmetrical-cut diagrams in the Appendix.

To obtain the double scattering contribution
to the semi-inclusive processes of hadron production in
Eq.~(\ref{factorize}), one will
then have to calculate the second derivatives of the
rescattering part $\overline{H}^D$.

After a closer examination of these rescattering parts,
one can find that all contributions from the
asymmetrical-cut diagrams have the form as
\begin{eqnarray}
\overline{H}^D_{asym}=\frac{ \vec{\ell_T} \cdot
(\vec{\ell_T}-f(z)\vec{k_T}) }
{\ell_T^2(\vec{\ell_T}-f(z)\vec{k_T})^2}
e^{iXp^+Y^-}, \label{eq:asym}
\end{eqnarray}
where $ f(z)=0,\,1,\,1-z,\,z$ is only a function of $z$,
 $X$ is the longitudinal momentum fraction and $Y^-$ the
spatial coordinates. One can prove that the second
derivative of the above expression vanishes at $k_T=0$,
\begin{eqnarray}
\nabla^2_{k_T}\frac{ \vec{\ell_T} \cdot (\vec{\ell_T}-f(z)\vec{k_T}) }
{\ell_T^2(\vec{\ell_T}-f(z)\vec{k_T})^2}  =0 \,.
\label{derivative}
\end{eqnarray}
Therefore, all contributions from the asymmetrical-cut(right-cut
and left-cut) diagrams will vanish after we make the second
partial derivative with respect to $k_T$ when we keep only the
leading terms up to ${\cal O}(x_B/Q^2\ell_T^2)$,
\begin{eqnarray}
\nabla^2_{k_T}\overline{H}^D_{asym}|_{k_T=0}
&=&0 +{\cal O}(x_B/Q^2\ell_T^2).
\label{ep:asym-deri}
\end{eqnarray}
In the same way we find that some of the central-cut diagrams will
not contribute to the final results, either.
In fact, after making the second partial derivative
with respect to $k_T$ only four central-cut
diagrams shown in Fig.~\ref{fig2} will contribute to the final result.

Including only those contributions that does not vanish after
second derivative with respect to $k_T$, we have

\begin{eqnarray}
\nabla^2_{k_T}\overline{H}^D|_{k_T=0}
&=&\int d\ell_T^2 \frac{\alpha_s}{2\pi} \frac{1+z^2}{1-z}
e^{i(x+x_L)p^+y^-} \frac{2\pi\alpha_s}{N_c}
\theta(-y_2^-)\theta(y^--y_1^-)
\nonumber \\
&\times&\left[
\frac{4C_A}{\ell_T^4}(1-e^{-ix_Lp^+y_2^-})
(1-e^{-ix_Lp^+(y^--y_1^-)}) \right. \nonumber  \\
&+&\frac{4C_F(1-z)^2}{\ell_T^4}
e^{-ix_Lp^+(y^--y_1^-)}e^{-ix_Lp^+y_2^-} \nonumber \\
&+&\frac{2C_A(1-z)}{\ell_T^4}e^{-ix_Lp^+y_2^-}
(1-e^{-ix_Lp^+(y^--y_1^-)})  \nonumber  \\
&+&\left. \frac{2C_A(1-z)}{\ell_T^4}e^{-ix_Lp^+(y^--y_1^-)}
(1-e^{-ix_Lp^+y_2^-}) +{\cal O}(x_B/Q^2\ell_T^2)\right]
\, \label{eq:expand1}
\end{eqnarray}
The first term at the right-hand side 
in Eq.~(\ref{eq:expand1}) comes from the
contribution of $ \overline{H}^D_{C(a)} $ which is the main contribution
in the previous calculation \cite{GW} with helicity amplitude approximation.
It contains hard-soft, double hard processes and their interferences.
The other three terms come from diagram (b),(c),(d) of
Fig.~\ref{fig2} respectively. They constitute corrections
to the first term in powers of $1-z$. The second term that are
proportional to $(1-z)^2$ is from the
final state radiation from the quark in the double hard
process in Fig.~\ref{fig2}-(b).
This term is the only contribution to the finite photon spectra
in the corresponding QED bremsstrahlung.
The third and fourth terms are the results of the interference of
the final state radiation from the quark and other radiation processes
(initial state radiation and radiation from the gluon line).
They contain both double
hard processes and interferences between hard-soft and double hard
processes in Fig.~\ref{fig2}-(c) and (d). In the limit of soft
gluon radiation, $(1-z)\rightarrow 0$,
these three terms can be neglected and
we recover the result in the
helicity amplitude approximation \cite{GW}.

\section{Modified Fragmentation Function and Parton Energy Loss}

Substituting Eq.~(\ref{eq:expand1}) into Eq.~(\ref{eq:hc0}),
Eq.~(\ref{factorize}) and adding the gluon fragmentation
processes, we have the semi-inclusive tensor from double
quark-gluon scattering including the contribution beyond the helicity
amplitude approximation,
\begin{eqnarray}
\frac{W_{\mu\nu}^{D,q}}{dz_h}
&=&\sum_q \,\int dx H^{(0)}_{\mu\nu}(xp,q)
\int_{z_h}^1\frac{dz}{z}D_{q\rightarrow h}(z_h/z)
\frac{\alpha_s}{2\pi} C_A \frac{1+z^2}{1-z} \nonumber \\
&\times&\int \frac{d\ell_T^2}{\ell_T^4} \frac{2\pi\alpha_s}{N_c}
\left[T^{A}_{qg}(x,x_L) +(1-z) T^{A(1)}_{qg}(x,x_L)
  +\frac{C_F}{C_A} (1-z)^2 T^{A(2)}_{qg}(x,x_L)\right] \nonumber \\
&+& (g-{\rm fragmentation})+({\rm virtual\,\, corrections})\, , \label{wd1}
\end{eqnarray}
where
\begin{eqnarray}
T^{A}_{qg}(x,x_L)&=& \int \frac{dy^{-}}{2\pi}\, dy_1^-dy_2^-
(1-e^{-ix_Lp^+y_2^-})(1-e^{-ix_Lp^+(y^--y_1^-)}) e^{i(x+x_L)p^+y^-}
\nonumber  \\
& &\frac{1}{2}\langle A | \bar{\psi}_q(0)\,
\gamma^+\, F_{\sigma}^{\ +}(y_{2}^{-})\, F^{+\sigma}(y_1^{-})\,\psi_q(y^{-})
| A\rangle \theta(-y_2^-)\theta(y^- -y_1^-) \;\; , \label{Tqg} \\
T^{A(1)}_{qg}(x,x_L)&=& \int \frac{dy^{-}}{2\pi}\, dy_1^-dy_2^-
\left[e^{-ix_Lp^+(y^- - y^-_1)}+e^{-ix_Lp^+y^-_2}
-2e^{-ix_Lp^+(y^--y^-_1+y_2^-)}\right]
\nonumber  \\
&& e^{i(x+x_L)p^+y^-}\frac{1}{4}\langle A | \bar{\psi}_q(0)\,
\gamma^+\, F_{\sigma}^{\ +}(y_{2}^{-})\, F^{+\sigma}(y_1^{-})\,\psi_q(y^{-})
| A\rangle \theta(-y_2^-)\theta(y^- -y_1^-) \;\; , \label{Tqg1} \\
T^{A(2)}_{qg}(x,x_L)&=& \int \frac{dy^{-}}{2\pi}\, dy_1^-dy_2^-
e^{ixp^+y^- +ix_Lp^+(y^-_1-y^-_2)}
\nonumber  \\
& & \frac{1}{2}\langle A | \bar{\psi}_q(0)\,
\gamma^+\, F_{\sigma}^{\ +}(y_{2}^{-})\, F^{+\sigma}(y_1^{-})\,\psi_q(y^{-})
| A\rangle \theta(-y_2^-)\theta(y^- -y_1^-)
\label{Tqg2}
\end{eqnarray}
are twist-four parton matrix elements of the nucleus. Apparently these
parton matrix elements are not independent of each other. $T^A_{qg}(x,x_L)$
has the complete four terms of soft-hard, double hard processes
and their interferences. Therefore it contains essentially four
independent parton matrix elements. $T^{A(1)}_{qg}(x,x_L)$ and
$T^{A(2)}_{gq}(x,x_L)$ are the results of the corrections beyond
helicity amplitude approximation. But these two matrix elements are
already contained in $T^A_{qg}(x,x_L)$.

During the collinear expansion, we have kept $\ell_T$ finite and
took the limit $k_T\rightarrow 0$. As a consequence, the gluon field
in one of the twist-four parton matrix elements in
Eqs.~(\ref{Tqg})-(\ref{Tqg2}) carries zero momentum in
the soft-hard process. However, the
gluon distribution $xf_g(x)$ at $x=0$ is not defined in QCD.
As argued in Ref.~\cite{GW}, this is due to the omission
of higher order terms in the collinear expansion.
As a remedy to the problem, a subset of the higher-twist
terms in the collinear expansion can be resummed to
restore the phase factors such as $\exp(ix_Tp^+y^-)$,
where $x_T\equiv \langle k_T^2\rangle/2p^+q^-z$ is related to
the intrinsic transverse momentum of the initial partons.
As a result, soft gluon fields in the parton matrix elements
will carry a fractional momentum $x_T$.

Using the factorization approximation \cite{GW,LQS,OW} we can
relate the twist-four parton matrix elements of the nucleus
to the twist-two parton distributions of nucleons and the
nucleus,
\begin{equation}
T^A_{qg}(x,x_L)=\frac{C}{x_A}
(1-e^{-x_L^2/x_A^2}) [f_q^A(x+x_L)\, x_Tf_g^N(x_T)
+f_q^A(x)(x_L+x_T)f_g^N(x_L+x_T)] \,
\end{equation}
where C is a constant, $x_A=1/MR_A$, $f_q^A(x)$ is the quark
distribution inside a nucleus, and $f_g^N(x)$ is the gluon
distribution inside a nucleon. A Gaussian distribution in the
light-cone coordinates was assumed for the nuclear distribution,
$\rho(y^-)=\rho_0 \exp({y^-}^2/2{R^-_A}^2)$, where
$R^-_A=\sqrt{2}R_AM/p^+$ and $M$ is the nucleon mass. We should
emphasize that the twist-four matrix elements is proportional to
$1/x_A=R_AM$, or the nuclear size \cite{OW}.

Notice that the off-diagonal matrix elements that correspond to
the interferences between hard-soft and double hard processes
is suppressed by a factor of $\exp(-x_L^2/x_A^2)$.
This is because in the interferences between double-hard and
hard-soft processes, there is actually
momentum flow of $x_Lp^+$ between the two nucleons where
the initial quark and gluon come from.
Without strong long range two-nucleon correlation inside a
nucleus, the amount of momentum flow $x_Lp^+$ should then
be restricted to the amount allowed by the uncertainty
principle, $1/R^-_A\sim p^+/R_AM$.
Similarly, the other two parton matrix elements in
Eqs.~(\ref{Tqg1}) and (\ref{Tqg2}) can be approximated as
\begin{eqnarray}
T^{A(1)}_{qg}(x,x_L)
&=&\frac{C}{2x_A}
\left\{ \left[f_q^A(x+x_L)\, x_Tf_g^N(x_T)
+f_q^A(x)(x_L+x_T)f_g^N(x_L+x_T)\right]e^{-x_L^2/x_A^2} \right. \nonumber\\
&-&\left. 2f_q^A(x)(x_L+x_T)f_g^N(x_L+x_T) \right\} \, , \\
T^{A(2)}_{qg}(x,x_L)
&=&\frac{C}{x_A}f_q^A(x)(x_L+x_T)f_g^N(x_L+x_T) \, .
\end{eqnarray}

From the above estimate of the matrix elements, both $T^A_{qg}(x,x_L)$
and $T^{A(1)}_{qg}(x,x_L)$ contain a factor $1-e^{-x_L^2/x_A^2}$
because of the LPM interference effect. Such an interference
factor will effectively cut off the integration over the
transverse momentum at $x_L\sim x_A$ in Eq.~(\ref{wd1}). As we
will show later in the calculation of the effective energy loss,
the integration with such a restriction in the transverse momentum
due to LPM interference effect will give rise to a factor $1/x_A$
in addition to the coefficient $f^A_q(x)/x_A$. Consequently,
contributions from double scattering in Eq.~(\ref{wd1}) that are
associated with $T^A_{qg}(x,x_L)$ and $T^{A(1)}_{qg}(x,x_L)$ will
be proportional to $R_A^2f^A_q(x)$. These are the leading double
scattering contributions in the limit of large nuclear size. On
the other hand, the third term $T^{A(2)}_{qg}(x,x_L)$ in
Eq.~(\ref{wd1}), which does not contain any interference effect,
will only contribute to a correction that is proportional to
$R_Af^A_q(x)$. In the limit of a large nucleus, $A^{1/3}\gg 1$, we
will neglect this term in the double scattering processes. It is
interesting to point out, however, that the physical process
associated with this term is totally responsible for the
non-vanishing photon spectra in QED bremsstrahlung which otherwise
vanishes in the helicity amplitude approximation. As we can see,
the leading correction beyond  helicity amplitude approximation
comes from the interference between this process and other
radiation processes that contribute to the leading result in the
first term.

The virtual correction in Eq.~(\ref{wd1}) can be obtained via
unitarity requirement similarly as in Ref.~\cite{GW}.
Including these virtual corrections and the single scattering
contribution, we can rewrite the semi-inclusive tensor in
terms of a modified fragmentation function
$\widetilde{D}_{q\rightarrow h}(z_h,\mu^2)$,
\begin{equation}
\frac{dW_{\mu\nu}}{dz_h}=\sum_q \int dx \widetilde{f}_q^A(x,\mu_I^2)
H^{(0)}_{\mu\nu}(x,p,q)
\widetilde{D}_{q\rightarrow h}(z_h,\mu^2) \label{eq:Wtot}
\end{equation}
where $\widetilde{f}_q^A(x,\mu_I^2)$ is the quark distribution functions
which in principle should also include the
higher-twist contribution \cite{MQiu} of the
initial state scattering. The modified effective quark
fragmentation function  is defined as
\begin{eqnarray}
\widetilde{D}_{q\rightarrow h}(z_h,\mu^2)&\equiv&
D_{q\rightarrow h}(z_h,\mu^2)
+\int_0^{\mu^2} \frac{d\ell_T^2}{\ell_T^2}
\frac{\alpha_s}{2\pi} \int_{z_h}^1 \frac{dz}{z}
\left[ \Delta\gamma_{q\rightarrow qg}(z,x,x_L,\ell_T^2)
D_{q\rightarrow h}(z_h/z) \right. \nonumber \\
&+& \left. \Delta\gamma_{q\rightarrow gq}(z,x,x_L,\ell_T^2)
D_{g\rightarrow h}(z_h/z)\right] \, , \label{eq:MDq}
\end{eqnarray}
where $D_{q\rightarrow h}(z_h,\mu^2)$ and
$D_{g\rightarrow h}(z_h,\mu^2)$ are the leading-twist
fragmentation functions. The modified splitting functions are
given as
\begin{eqnarray}
\Delta\gamma_{q\rightarrow qg}(z,x,x_L,\ell_T^2)&=&
\left[\frac{1+z^2}{(1-z)_+}T^{A(m)}_{qg}(x,x_L) +
\delta(1-z)\Delta T^{A(m)}_{qg}(x,\ell_T^2) \right]
\frac{2\pi\alpha_s C_A}
{\ell_T^2 N_c\widetilde{f}_q^A(x,\mu_I^2)}\, ,
\label{eq:r1}\\
\Delta\gamma_{q\rightarrow gq}(z,x,x_L,\ell_T^2)
&=& \Delta\gamma_{q\rightarrow qg}(1-z,x,x_L,\ell_T^2) \label{eq:r2}, \\
\Delta T^{A(m)}_{qg}(x,\ell_T^2) &\equiv &
\int_0^1 dz\frac{1}{1-z}\left[ 2 T^{A(m)}_{qg}(x,x_L)|_{z=1}
-(1+z^2) T^{A(m)}_{qg}(x,x_L)\right] \, , \label{eq:delta-T} \\
T^{A(m)}_{qg}(x,x_L) &\equiv & T^{A}_{qg}(x,x_L)
+(1-z)T^{A(1)}_{qg}(x,x_L) \, . \label{modT}
\end{eqnarray}

The above modified fragmentation function is almost the same as in the
previous calculation with helicity amplitude approximation, except that
the twist-four parton matrix element $T^{A}_{qg}(x,x_L)$ is replaced by
a modified one $T^{A(m)}_{qg}(x,x_L)$ in Eq.~(\ref{modT}). One can then
calculate numerically the modified fragmentation function as in
Refs.~\cite{EW1,GW}. To further simplify the calculation, we assume
$x_T\ll x_L \ll x$. The modified parton matrix elements can be
approximated by
\begin{equation}
T^{A(m)}_{qg}(x,x_L)\approx \frac{\widetilde{C}}{x_A}
(1-e^{-x_L^2/x_A^2}) f_q^A(x) \left[1-\frac{1-z}{2}\right],
\label{modT2}
\end{equation}
where $\widetilde{C}\equiv 2C x_Tf^N_g(x_T)$ is a coefficient which
should in principle depends on $Q^2$ and $x_T$. Here we will simply take
it as a constant. The new correction term
in this calculation is thus negative in the modified splitting function.
This will reduce the nuclear suppression of hadron spectra at large values
of $z$ and thus reduce the effective quark energy loss.

\begin{figure}
\centerline{\psfig{file=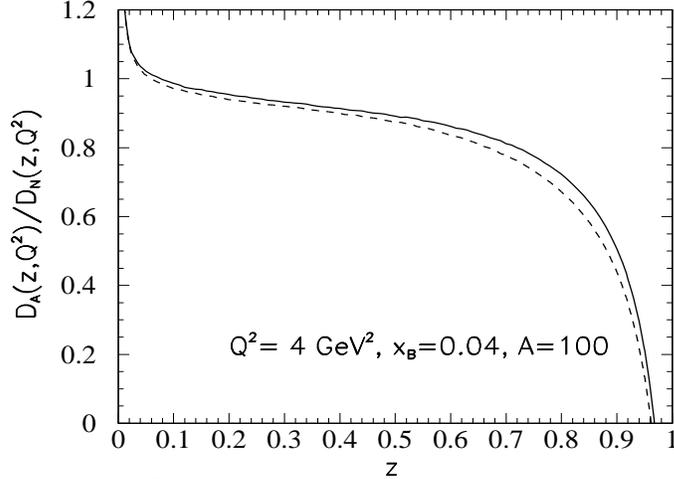,width=3.5in,height=2.5in}}
\caption{Calculated nuclear modification factor for the quark
fragmentation in a nucleus (A=100). The solid line is the current
calculation with the new correction term. The dashed line is
the previous result with helicity amplitude approximation.}
\label{fig3}
\end{figure}

Because of momentum conservation, the fractional momentum in a
nucleon is limited to $x_L<1$. Though the Fermi motion effect in a
nucleus can allow $x_L>1$, the parton distribution in this region
is still significant suppressed. It therefore provides a natural
cut-off for $x_L$ in the integration over $z$ and $\ell_T$ in 
Eq.~(\ref{eq:MDq}). Shown
in Fig.~\ref{fig3} are the calculated nuclear modification factor
for the quark fragmentation function $D_A(z,Q^2)/D_N(z,Q^2)$
inside  a nucleus with $A=100$. In this numerical evaluation, we
have taken $\widetilde{C}=0.006$ GeV$^2$ which was fitted to the
HERMES experimental data \cite{EW1}. The dashed curve is for the
modified fragmentation function in the helicity amplitude
approximation and the solid curve is obtained with the new
correction term. Apparently, the new correction term reduces the
nuclear modification, though the reduction is not very
significant.

Similarly, we can also calculate the effective quark energy loss,
which is defined as the energy carried away by the radiated gluon,
\begin{eqnarray}
\langle\Delta z_g\rangle(x_B,\mu^2)
&=& \int_0^{\mu^2}\frac{d\ell_T^2}{\ell_T^2}
\int_0^1 dz \frac{\alpha_s}{2\pi}
 z\,\Delta\gamma_{q\rightarrow gq}(z,x_B,x_L,\ell_T^2) \nonumber \\
&=&\frac{C_A\alpha_s^2}{N_c} \int_0^{\mu^2}\frac{d\ell_T^2}{\ell_T^4}
\int_0^1 dz [1+(1-z)^2]
\frac{T^{A(m)}_{qg}(x_B,x_L)}{\widetilde{f}_q^A(x_B,\mu_I^2)}\, .
\label{eq:loss1}
\end{eqnarray}
We separate the parton energy loss as two parts
\begin{eqnarray}
\langle\Delta z_g\rangle(x_B,\mu^2)
=\langle\Delta z_g\rangle_{heli}(x_B,\mu^2)
  + \langle\Delta z_g\rangle_{corr}(x_B,\mu^2),
\label{eq:all-loss}
\end{eqnarray}
where $\langle\Delta
z_g\rangle_{heli}(x_B,\mu^2)$ is the leading quark energy loss with
helicity amplitude approximation \cite{GW}, and $\langle\Delta
z_g\rangle_{corr}(x_B,\mu^2)$ is the new correction to the quark energy loss
in this calculation.  Using the approximation for the
modified twist-four parton matrix elements
in Eq.~(\ref{modT2}), we have
\begin{eqnarray}
\langle\Delta z_g\rangle_{heli}(x_B,\mu^2)
&=&\widetilde{C}\frac{C_A\alpha_s^2}{N_c}
\frac{x_B}{x_AQ^2} \int_0^1 dz \frac{1+(1-z)^2}{z(1-z)}
\int_0^{x_\mu} \frac{dx_L}{x_L^2} (1-e^{-x_L^2/x_A^2});
\label{eq:heli-loss}  \\
\langle\Delta z_g\rangle_{corr}(x_B,\mu^2)
&=&\widetilde{C}\frac{C_A\alpha_s^2}{N_c}
\frac{x_B}{x_AQ^2} \int_0^1 dz \frac{1+(1-z)^2}{z(1-z)}
\int_0^{x_\mu} \frac{dx_L}{x_L^2}
(-\frac{z }{2}) (1-e^{-x_L^2/x_A^2}) \, ,  \label{eq:corr1-loss} \\
\end{eqnarray}
where $x_\mu=\mu^2/2p^+q^-z(1-z)=x_B/z(1-z)$ if we choose the
factorization scale as $\mu^2=Q^2$.
When $x_A\ll x_B\ll 1$ we can estimate the leading quark energy
loss roughly as
\begin{eqnarray}
\langle \Delta z_g\rangle_{heli}(x_B,\mu^2)& \approx &
\widetilde{C}\frac{C_A\alpha_s^2}{N_c}\frac{x_B}{Q^2
x_A^2}2\sqrt{\pi}[3\ln\frac{1-2x_B}{x_B}-1] \, ,
\label{eq:appr1-loss} \\
\langle \Delta z_g\rangle_{corr}(x_B,\mu^2)& \approx &
-\widetilde{C}\frac{C_A\alpha_s^2}{N_c}\frac{x_B}{Q^2
x_A^2}\sqrt{\pi}[\ln\frac{1-2x_B}{x_B}+\frac{1}{2}] \, .
\label{eq:appr2-loss}
\end{eqnarray}
Since $x_A=1/MR_A$, both of the energy loss $\langle \Delta
z_g\rangle_{heli}$ and $ \langle \Delta z_g\rangle_{corr}$ depend
quadratically on the nuclear size. Adding them together, we have
\begin{eqnarray}
\langle\Delta z_g\rangle(x_B,\mu^2) & \approx &
\frac{\tilde{C}\alpha_s^2}{N_c}\frac{x_B}{Q^2
x_A^2}\sqrt{\pi}[5\ln\frac{1-2x_B}{x_B}-\frac{5}{2}]  \label{all-loss2} \\
& \approx & \frac{5}{6}\langle \Delta z_g\rangle_{heli}(x_B,\mu^2)
,\  \  \  (x_A\ll x_B\ll 1 ) \, .
\end{eqnarray}
The new correction thus reduces the effective quark energy loss by
approximately a factor of $5/6$ from the result with helicity
amplitude approximation.

\section{Summary}

We have extended an earlier study \cite{GW} on gluon radiation
induced by multiple parton scattering in DIS off a nuclear target
with a complete calculation beyond the helicity amplitude (or soft radiation)
approximation. Working within the framework of the generalized
factorization of twist-four processes, we obtained a new
correction to the modified parton fragmentation functions. Such a
new correction essentially results in a new term in the modified
splitting function which is proportional to $(1-z)$. In the limit
of helicity amplitude approximation $(1-z)\rightarrow 0$, this
term vanishes and we recover the early results \cite{GW}.

The new correction we obtained in this paper comes from the gluon
radiation process (residual final state radiation from a quark
after incomplete cancellation by the initial state radiation) that
is actually responsible for the photon radiation in QED. However,
the leading correction beyond the helicity amplitude approximation
does not come from this process itself. Rather, it comes from the
interference between this process and the other gluon radiation
processes that are responsible for the result in the helicity
amplitude approximation. Though it is not dominant for induced
gluon radiation in QCD, it still make a finite contribution to the
modified fragmentation function for a quark propagating inside a
nuclear medium and to the effective quark energy loss. We found
that it reduces the effective quark energy loss by a factor of
$5/6$.

\section*{Acknowledgements}
We would like to thank Enke Wang for numerous discussions throughout
this work. This work was supported by NSFC under project Nos. 19928511 and
10135030, and by the Director, Office of Energy Research, Office of
High Energy and Nuclear Physics, Divisions of Nuclear Physics, of the U.S.
Department of Energy under Contract No. DE-AC03-76SF00098.

\section*{Appendix}
\begin{figure}
\centerline{\psfig{file=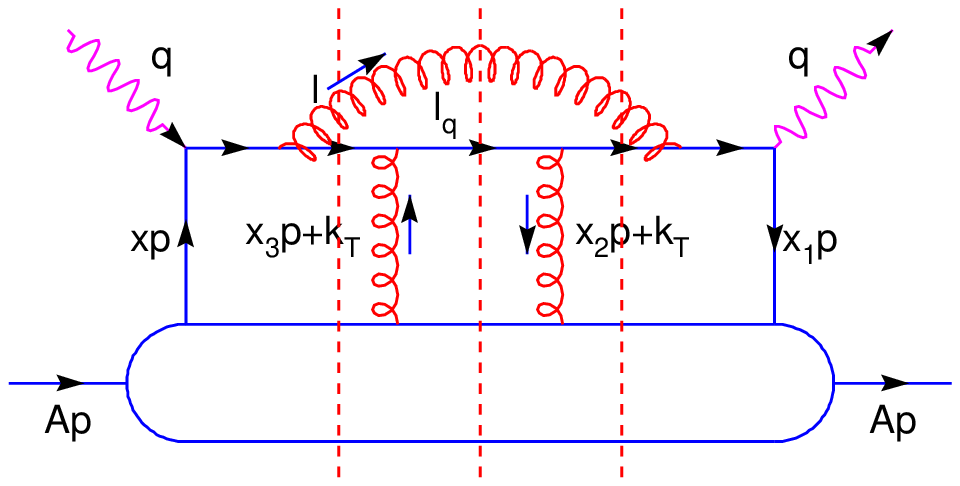,width=2.6in,height=1.8in}}
\caption{   }
\label{Ap-1}
\end{figure}

\begin{figure}
\centerline{\psfig{file=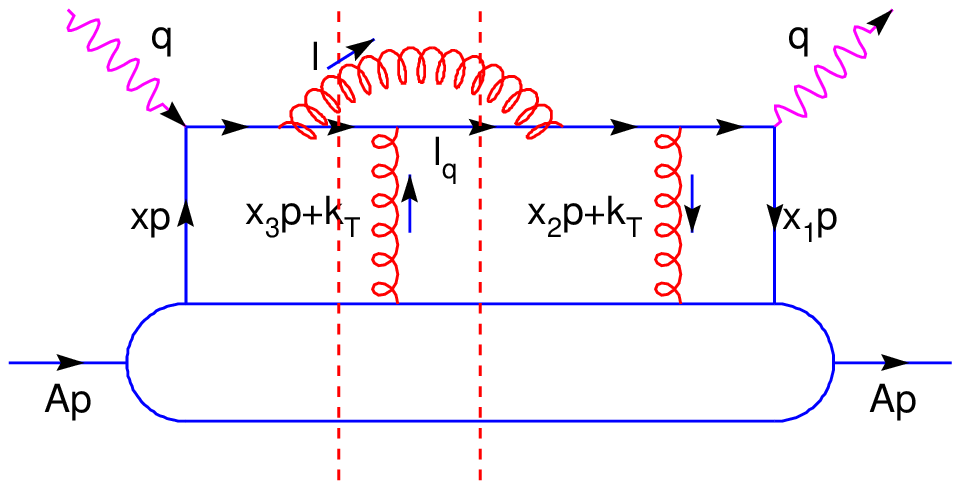,width=2.6in,height=1.8in}}
\caption{ }
\label{Ap-2}
\end{figure}

\begin{figure}
\centerline{\psfig{file=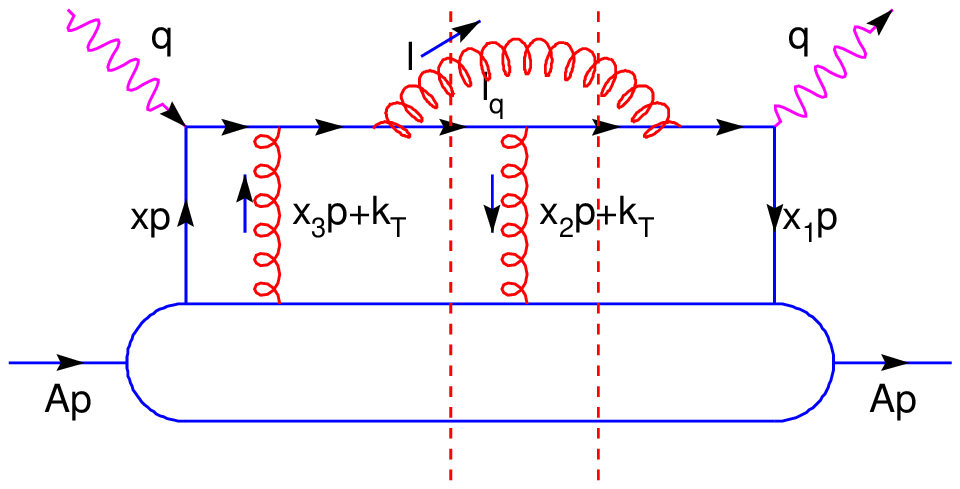,width=2.6in,height=1.8in}}
\caption{ }
\label{Ap-3}
\end{figure}

\begin{figure}
\centerline{\psfig{file=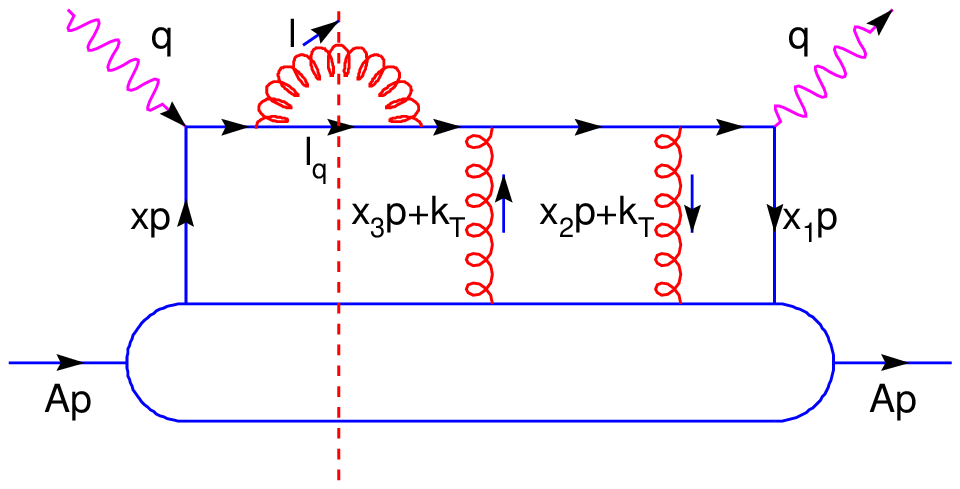,width=2.6in,height=1.8in}}
\caption{}
\label{Ap-4}
\end{figure}

\begin{figure}
\centerline{\psfig{file=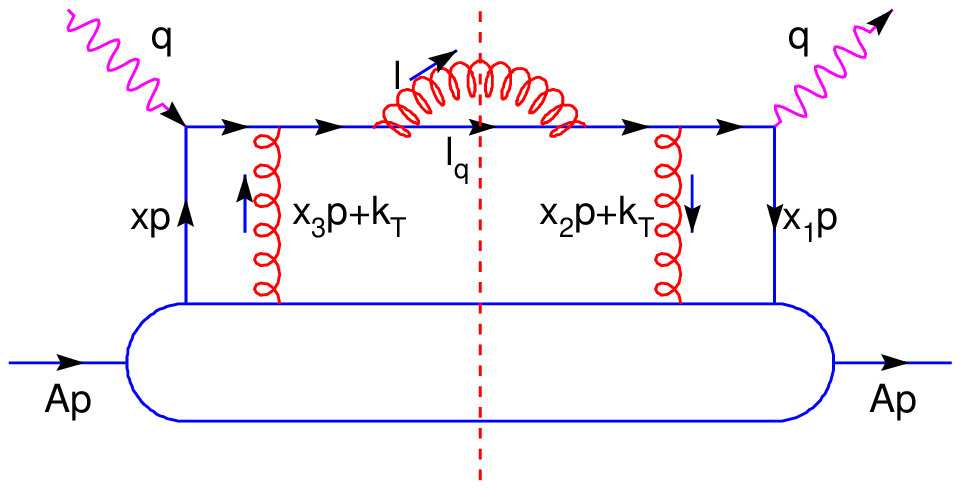,width=2.6in,height=1.8in}}
\caption{   }
\label{Ap-5}
\end{figure}

\begin{figure}
\centerline{\psfig{file=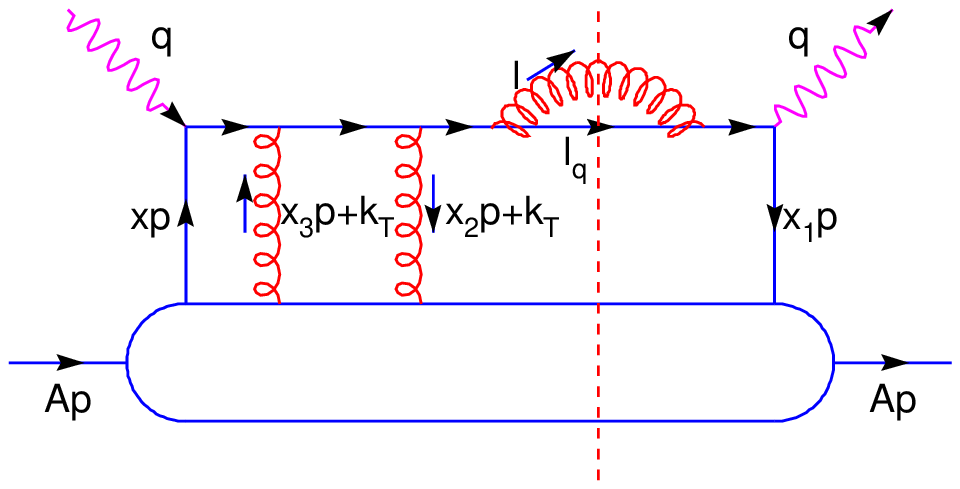,width=2.6in,height=1.8in}}
\caption{ }
\label{Ap-6}
\end{figure}

\begin{figure}
\centerline{\psfig{file=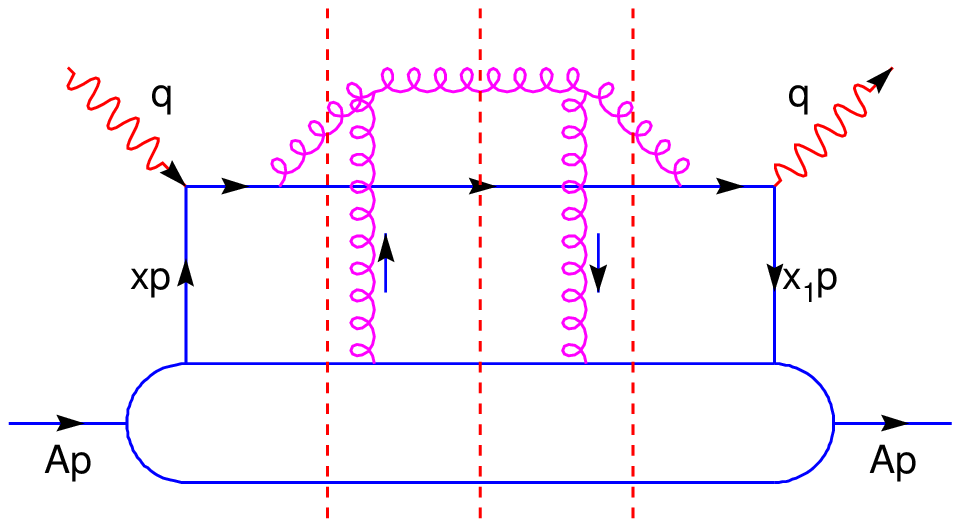,width=2.6in,height=1.8in}}
\caption{   }
\label{Ap-7}
\end{figure}

\begin{figure}
\centerline{\psfig{file=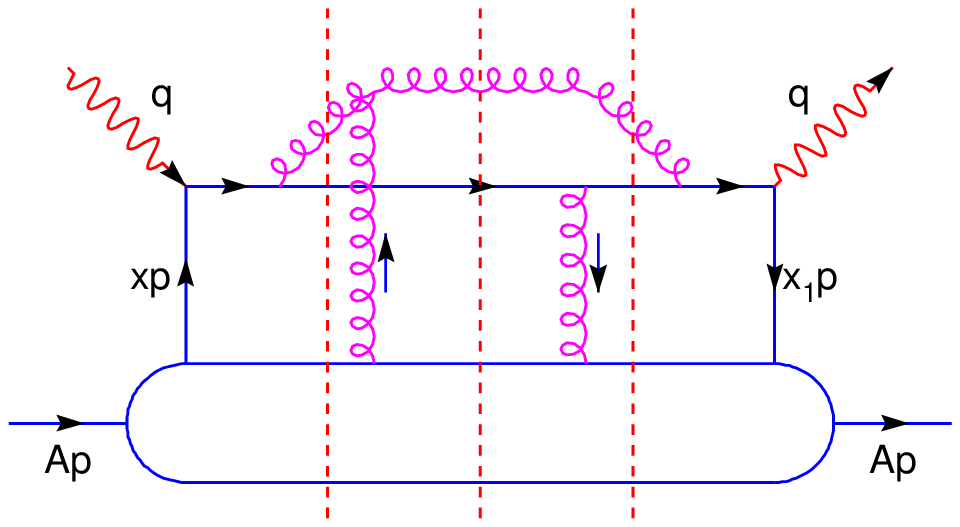,width=2.6in,height=1.8in}}
\caption{   }
\label{Ap-8}
\end{figure}

\begin{figure}
\centerline{\psfig{file=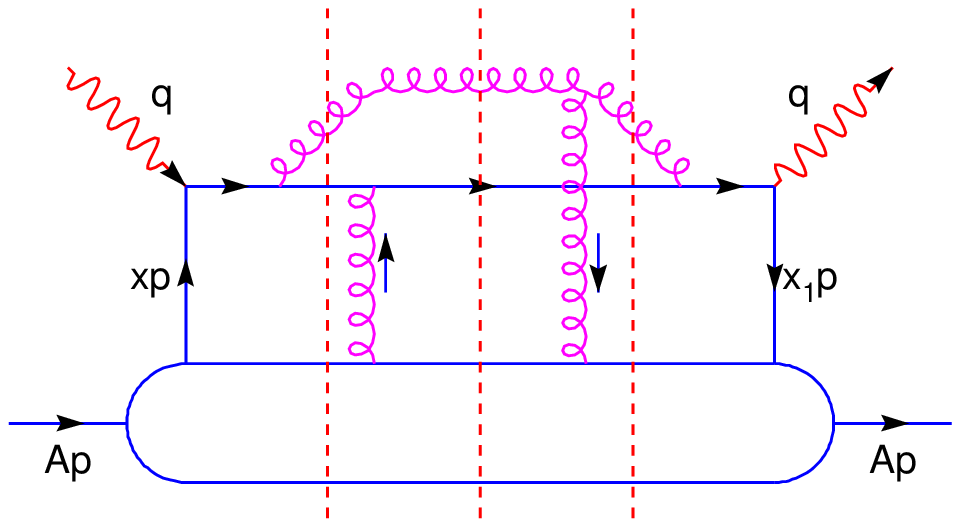,width=2.6in,height=1.8in}}
\caption{   }
\label{Ap-9}
\end{figure}

\begin{figure}
\centerline{\psfig{file=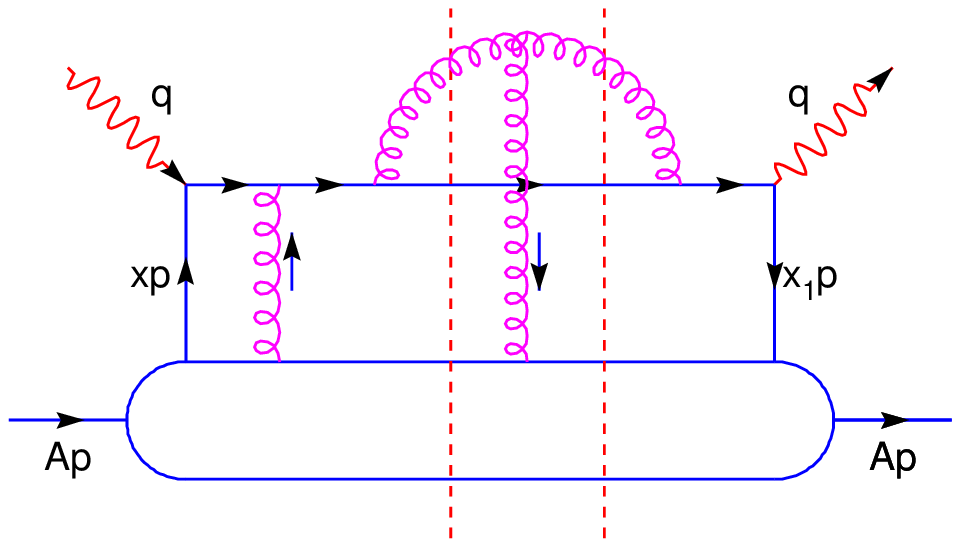,width=2.6in,height=1.8in}}
\caption{   }
\label{Ap-10}
\end{figure}

\begin{figure}
\centerline{\psfig{file=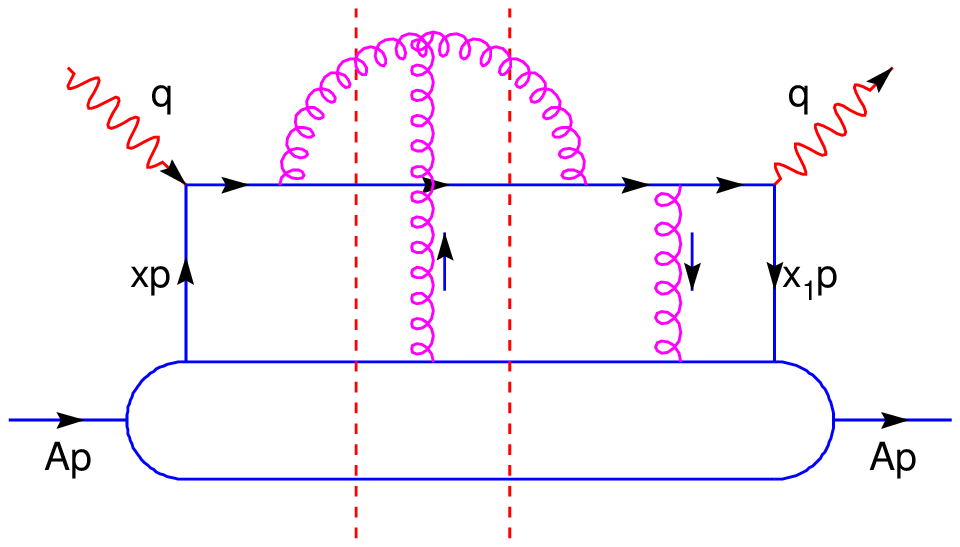,width=2.6in,height=1.8in}}
\caption{   }
\label{Ap-11}
\end{figure}

In this Appendix we will list our complete calculation of  quark-gluon
double scattering in detail. There are total 23 cut diagrams which are
illustrated in Figs.~\ref{Ap-1}-\ref{Ap-11}.


In Fig.~\ref{Ap-1} there are three possible cuts(central cut,
left cut and right cut). We get
\begin{eqnarray}
\overline{H}^D_{Ap-1}(y^-,y_1^-,y_2^-,k_T,x,p,q,z)&=&
\int \frac{d\ell_T^2}{\ell_T^2}\, \frac{\alpha_s}{2\pi}C_F\,
 \frac{1+z^2}{1-z} \nonumber \\
&\times&\frac{2\pi\alpha_s}{N_c}
\overline{I}_{Ap-1}(y^-,y_1^-,y_2^-,\ell_T,k_T,x,p,q,z)
 \, , \label{eq:Ap-1}
\end{eqnarray}
where
\begin{eqnarray}
\overline{I}_{Ap-1,C}(y^-,y_1^-,y_2^-,\ell_T,k_T,x,p,q,z)
&=&e^{i(x+x_L)p^+y^- + ix_Dp^+(y_1^- - y_2^-)}
\theta(-y_2^-)\theta(y^- - y_1^-) \nonumber \\
&\times &(1-e^{-ix_Lp^+y_2^-})(1-e^{-ix_Lp^+(y^- - y_1^-)}) \; .
\label{eq:I1C}  \\
\overline{I}_{Ap-1,L}(y^-,y_1^-,y_2^-,\ell_T,k_T,x,p,q,z)\,
&=&-e^{i(x+x_L)p^+y^- + ix_Dp^+(y_1^- - y_2^-)}
\theta(y_1^- - y_2^-)\theta(y^- - y_1^-) \nonumber \\
&\times &(1-e^{-ix_Lp^+(y^- - y_1^-)}) \, ,\label{eq:I1L} \\
\overline{I}_{Ap-1,R}(y^-,y_1^-,y_2^-,\ell_T,k_T,x,p,q,z)
&=&-e^{i(x+x_L)p^+y^- + ix_Dp^+(y_1^- - y_2^-)}
\theta(-y_2^-)\theta(y_2^- - y_1^-) \nonumber \\
&\times &(1-e^{-ix_Lp^+y_2^-}) \, .\label{eq:I1R}
\end{eqnarray}

In Fig.~\ref{Ap-2}, there are two different cuts, central or left. So we
obtain,

\begin{eqnarray}
\overline{H}^D_{Ap-2}(y^-,y_1^-,y_2^-,k_T,x,p,q,z)&=&
\int d\ell_T^2 \frac{\vec{\ell_T} \cdot (\vec{\ell_T}-(1-z)\vec{k_T})}
 {\ell_T^2(\vec{\ell_T}-(1-z)\vec{k_T})^2 }\, \frac{\alpha_s}{2\pi}\,
(C_F-\frac{C_A}{2})\frac{1+z^2}{1-z} \nonumber \\
&\times&\frac{2\pi\alpha_s}{N_c}
\overline{I}_{Ap-2}(y^-,y_1^-,y_2^-,\ell_T,k_T,x,p,q,z)
 \, , \label{eq:hc2} \\
\overline{I}_{Ap-2,C}(y^-,y_1^-,y_2^-,\ell_T,k_T,x,p,q,z)&=&
e^{i(x+x_L)p^+y^-+ix_Dp^+(y_1^--y_2^-)}
\theta(-y_2^-)\theta(y^- - y_1^-) \nonumber \\
&\times&[e^{-ix_Lp^+(y^- - y_1^-)}-e^{-ix_Lp^+(y^- - y_1^-+y_2^-)}] \, ,
\label{eq:I2C} \\
\overline{I}_{Ap-2,L}(y^-,y_1^-,y_2^-,\ell_T,k_T,x,p,q,z)&=&
e^{i(x+x_L)p^+y^-+ix_Dp^+(y_1^--y_2^-)}
\theta(y^- - y_1^-)\theta(y_1^- - y_2^-) \nonumber \\
&\times&[e^{-ix_Lp^+(y^- - y_2^-)+i(x_D^0-x_D)p^+(y_1^- - y_2^-)}
 -e^{-ix_Lp^+(y^- - y_1^-)}] \, ,
\label{eq:I2L}
\end{eqnarray}
where $x_D^0=k_T/2p^+q^-$.

As for the central cut and right cut of Fig.~\ref{Ap-3}, we obtain
\begin{eqnarray}
\overline{H}^D_{Ap-3}(y^-,y_1^-,y_2^-,k_T,x,p,q,z)&=&
\int d\ell_T^2 \frac{\vec{\ell_T} \cdot (\vec{\ell_T}-(1-z)\vec{k_T})}
 {\ell_T^2(\vec{\ell_T}-(1-z)\vec{k_T})^2 }\, \frac{\alpha_s}{2\pi}\,
(C_F-\frac{C_A}{2}) \frac{1+z^2}{1-z} \nonumber \\
&\times&\frac{2\pi\alpha_s}{N_c}
\overline{I}_{Ap-3}(y^-,y_1^-,y_2^-,\ell_T,k_T,x,p,q,z)
 \, , \label{eq:hc3} \\
\overline{I}_{Ap-3,C}(y^-,y_1^-,y_2^-,\ell_T,k_T,x,p,q,z)&=&
e^{i(x+x_L)p^+y^-+ix_Dp^+(y_1^--y_2^-)}
\theta(-y_2^-)\theta(y^- - y_1^-) \nonumber \\
&\times&[e^{-ix_Lp^+y_2^-}-e^{-ix_Lp^+(y^- - y_1^-+y_2^-)}] \, ,
\label{eq:I3C} \\
\overline{I}_{Ap-3,R}(y^-,y_1^-,y_2^-,\ell_T,k_T,x,p,q,z)&=&
e^{i(x+x_L)p^+y^-+ix_Dp^+(y_1^--y_2^-)}
\theta(-y_2^-)\theta(y_2^- - y_1^-) \nonumber \\
&\times&[e^{-i(x_D^0-x_D)p^+(y_1^- - y_2^-)-ix_Lp^+y_1^-}
 -e^{-ix_Lp^+y_2^-}] \, .
\label{eq:I3R}
\end{eqnarray}

There is only one cut (left cut) in Fig.~\ref{Ap-4} with the
contribution,
\begin{eqnarray}
\overline{H}^D_{Ap-4}(y^-,y_1^-,y_2^-,k_T,x,p,q,z)&=&
\int \frac{d\ell_T^2}{\ell_T^2}\, \frac{\alpha_s}{2\pi}\,
C_F \frac{1+z^2}{1-z}  \nonumber \\
&\times&\frac{2\pi\alpha_s}{N_c}
\overline{I}_{Ap-4}(y^-,y_1^-,y_2^-,\ell_T,k_T,x,p,q,z)
 \, , \label{eq:hc} \\
\overline{I}_{Ap-4,L}(y^-,y_1^-,y_2^-,\ell_T,k_T,x,p,q,z)&=&
-e^{i(x+x_L)p^+y^-+ix_Dp^+(y_1^--y_2^-)}
\theta(y^- - y_1^-)\theta(y_1^- - y_2^-) \nonumber \\
&\times&e^{i(x_D^0-x_D)p^+(y_1^- - y_2^-)} e^{-ix_Lp^+(y^- - y_2^-)} \, .
\label{eq:I4L}
\end{eqnarray}

As for Fig.~\ref{Ap-5}, we get
\begin{eqnarray}
\overline{H}^D_{Ap-5}(y^-,y_1^-,y_2^-,k_T,x,p,q,z)&=&
\int \frac{d\ell_T^2}{(\vec{\ell_T}-(1-z)\vec{k_T})^2}\,
\frac{\alpha_s}{2\pi}\,
C_F \frac{1+z^2}{1-z}  \nonumber \\
&\times&\frac{2\pi\alpha_s}{N_c}
\overline{I}_{Ap-5}(y^-,y_1^-,y_2^-,\ell_T,k_T,x,p,q,z)
 \, , \label{eq:hc5} \\
\overline{I}_{Ap-5,C}(y^-,y_1^-,y_2^-,\ell_T,k_T,x,p,q,z)&=&
e^{i(x+x_L)p^+y^-+ix_Dp^+(y_1^--y_2^-)}
\theta(-y_2^-)\theta(y^- - y_1^-) \nonumber \\
&\times& e^{-ix_Lp^+(y^- - y_1^-)}e^{-ix_Lp^+y_2^-} \, .
\label{eq:I5C}
\end{eqnarray}

The contribution from Fig.~\ref{Ap-6} is
\begin{eqnarray}
\overline{H}^D_{Ap-6}(y^-,y_1^-,y_2^-,k_T,x,p,q,z)&=&
\int \frac{d\ell_T^2}{\ell_T^2}\, \frac{\alpha_s}{2\pi}\,
C_F \frac{1+z^2}{1-z}  \nonumber \\
&\times&\frac{2\pi\alpha_s}{N_c}
\overline{I}_{Ap-6}(y^-,y_1^-,y_2^-,\ell_T,k_T,x,p,q,z)
 \, , \label{eq:hc6}  \\
\overline{I}_{Ap-6,R}(y^-,y_1^-,y_2^-,\ell_T,k_T,x,p,q,z)&=&
-e^{i(x+x_L)p^+y^-+ix_Dp^+(y_1^--y_2^-)}
\theta(-y_2^-)\theta(y_2^- - y_1^-) \nonumber \\
&\times&e^{i(x_D^0-x_D)p^+(y_1^- - y_2^-)}e^{-ix_Lp^+y_1^-}
 \, .
\label{eq:I6R}
\end{eqnarray}


As for the processes in Fig.~\ref{Ap-7} we have three possible cuts.
Thus the contributions can be written as

\begin{eqnarray}
\overline{H}^D_{Ap-7,C}(y^-,y_1^-,y_2^-,k_T,x,p,q,z)&=&
\int \frac{d\ell_T^2}{(\vec{\ell_T}-\vec{k_T})^2}\, \frac{\alpha_s}{2\pi}\,
 C_A\frac{1+z^2}{1-z} \nonumber \\
&\times&\frac{2\pi\alpha_s}{N_c}
\overline{I}_{Ap-7,C}(y^-,y_1^-,y_2^-,\ell_T,k_T,x,p,q,z)
 \, , \label{eq:hc7-1} \\
\overline{I}_{Ap-7,C}(y^-,y_1^-,y_2^-,\ell_T,k_T,x,p,q,z)&=&
e^{i(x+x_L)p^+y^-+ix_Dp^+(y_1^--y_2^-)}
\theta(-y_2^-)\theta(y^- - y_1^-) \nonumber \\
&\times&[e^{ix_Dp^+y_2^-/(1-z)}-e^{-ix_Lp^+y_2^-}] \nonumber \\
&\times&[e^{ix_Dp^+(y^- - y_1^-)/(1-z)}-e^{-ix_Lp^+(y^- - y_1^-)}] \, ,
\label{eq:I7C}  \\
\overline{H}^D_{Ap-7,L(R)}(y^-,y_1^-,y_2^-,k_T,x,p,q,z)&=&
\int \frac{d\ell_T^2}{\ell_T^2}\, \frac{\alpha_s}{2\pi}\,
 C_A\frac{1+z^2}{1-z} \nonumber \\
&\times&\frac{2\pi\alpha_s}{N_c}
\overline{I}_{Ap-7,L(R)}(y^-,y_1^-,y_2^-,\ell_T,k_T,x,p,q,z)
 \, , \label{eq:hc7-2} \\
\overline{I}_{Ap-7,L}(y^-,y_1^-,y_2^-,\ell_T,k_T,x,p,q,z)&=&
-e^{i(x+x_L)p^+y^-+ix_Dp^+(y_1^--y_2^-)}
\theta(y^- - y_1^-)\theta(y_1^- - y_2^-) \nonumber \\
&\times& e^{-i(1-z/(1-z) )x_Dp^+(y_1^- - y_2^-)} \nonumber \\
&\times&[1-e^{-ix_Lp^+(y^- - y_1^-)}] \, ,
\label{eq:I7L}  \\
\overline{I}_{Ap-7,R}(y^-,y_1^-,y_2^-,\ell_T,k_T,x,p,q,z)&=&
-e^{i(x+x_L)p^+y^-+ix_Dp^+(y_1^--y_2^-)}
\theta(-y_2^-)\theta(y_2^- - y_1^-) \nonumber \\
&\times& e^{-i(1-z/(1-z) )x_Dp^+(y_1^- - y_2^-)}
[1-e^{-ix_Lp^+y_2^-}] \, ,
\label{eq:I7R}
\end{eqnarray}

There are three cuts in Fig.~\ref{Ap-8} and we have

\begin{eqnarray}
\overline{H}^D_{Ap-8}(y^-,y_1^-,y_2^-,k_T,x,p,q,z)&=&
\int d\ell_T^2 \frac{\vec{\ell_T} \cdot (\vec{\ell_T}-\vec{k_T}) }
 {\ell_T^2(\vec{\ell_T}-\vec{k_T})^2}\, \frac{\alpha_s}{2\pi}\,
\frac{C_A}{2}\frac{1+z^2}{1-z} \nonumber \\
&\times&\frac{2\pi\alpha_s}{N_c}
\overline{I}_{Ap-8}(y^-,y_1^-,y_2^-,\ell_T,k_T,x,p,q,z)
 \, , \label{eq:hc8} \\
\overline{I}_{Ap-8,C}(y^-,y_1^-,y_2^-,\ell_T,k_T,x,p,q,z)&=&
-e^{i(x+x_L)p^+y^-+ix_Dp^+(y_1^--y_2^-)}
\theta(-y_2^-)\theta(y^- - y_1^-) \nonumber \\
&\times&[e^{ix_Dp^+y_2^-/(1-z)}-e^{-ix_Lp^+y_2^-}] \nonumber \\
&\times&[1-e^{-ix_Lp^+(y^- - y_1^-)}] \, ,
\label{eq:I8C}  \\
\overline{I}_{Ap-8,L}(y^-,y_1^-,y_2^-,\ell_T,k_T,x,p,q,z)&=&
-e^{i(x+x_L)p^+y^-+ix_Dp^+(y_1^--y_2^-)}
\theta(y^- - y_1^-)\theta(y_1^- - y_2^-) \nonumber \\
&\times& e^{-i(1-z/(1-z))x_Dp^+(y_1^- - y_2^-)} \nonumber \\
&\times&[e^{-ix_Lp^+(y^- - y_1^-)} -e^{ix_Dp^+(y^- - y_1^-)/(1-z)}] \, ,
\label{eq:I8L}  \\
\overline{I}_{Ap-8,R}(y^-,y_1^-,y_2^-,\ell_T,k_T,x,p,q,z)&=&
-e^{i(x+x_L)p^+y^-+ix_Dp^+(y_1^--y_2^-)}
\theta(-y_2^-)\theta(y_2^- - y_1^-) \nonumber \\
&\times&[e^{-ix_Lp^+y_2^-} -e^{ix_Dp^+y_2^-/(1-z)}] \, ,
\label{eq:I8R}
\end{eqnarray}

The contributions from the three cuts in Fig.~\ref{Ap-9} can be read as

\begin{eqnarray}
\overline{H}^D_{Ap-9}(y^-,y_1^-,y_2^-,k_T,x,p,q,z)&=&
\int d\ell_T^2 \frac{\vec{\ell_T} \cdot (\vec{\ell_T}-\vec{k_T}) }
 {\ell_T^2(\vec{\ell_T}-\vec{k_T})^2}\, \frac{\alpha_s}{2\pi}\,
\frac{C_A}{2}\frac{1+z^2}{1-z} \nonumber \\
&\times&\frac{2\pi\alpha_s}{N_c}
\overline{I}_{Ap-9}(y^-,y_1^-,y_2^-,\ell_T,k_T,x,p,q,z)
 \, , \label{eq:hc9} \\
\overline{I}_{Ap-9,C}(y^-,y_1^-,y_2^-,\ell_T,k_T,x,p,q,z)&=&
-e^{i(x+x_L)p^+y^-+ix_Dp^+(y_1^--y_2^-)}
\theta(-y_2^-)\theta(y^- - y_1^-) \nonumber \\
&\times&[e^{ix_Dp^+(y^- - y_1^-)/(1-z)}-e^{-ix_Lp^+(y^- - y_1^-)}]
\nonumber \\
&\times&[1-e^{-ix_Lp^+ y_2^-}] \, ,
\label{eq:I9C}  \\
\overline{I}_{Ap-9,L}(y^-,y_1^-,y_2^-,\ell_T,k_T,x,p,q,z)&=&
-e^{i(x+x_L)p^+y^-+ix_Dp^+(y_1^--y_2^-)}
\theta(y^- - y_1^-)\theta(y_1^- - y_2^-) \nonumber \\
&\times&[e^{-ix_Lp^+ (y^- - y_1^-)} -e^{ix_Dp^+(y^- - y_1^-)/(1-z)}] \, ,
\label{eq:I9L}   \\
\overline{I}_{Ap-9,R}(y^-,y_1^-,y_2^-,\ell_T,k_T,x,p,q,z)&=&
-e^{i(x+x_L)p^+y^-+ix_Dp^+(y_1^--y_2^-)}
\theta(-y_2^-)\theta(y_2^- - y_1^-) \nonumber \\
&\times& e^{-i(1-z/(1-z))x_Dp^+(y_1^- - y_2^-)} \nonumber \\
&\times&[e^{-ix_Lp^+y_2^-} -e^{ix_Dp^+y_2^-/(1-z)}] \, ,
\label{eq:I9R}
\end{eqnarray}

In Fig.~\ref{Ap-10} there are two possible cuts (central or left). We have
\begin{eqnarray}
\overline{H}^D_{Ap-10,C}(y^-,y_1^-,y_2^-,k_T,x,p,q,z)&=&
\int d\ell_T^2\frac{ (\vec{\ell_T}-\vec{k_T})\cdot
(\vec{\ell_T}-(1-z)\vec{k_T}) }
{(\vec{\ell_T}-\vec{k_T})^2 (\vec{\ell_T}-(1-z)\vec{k_T})^2}\,
\frac{\alpha_s}{2\pi}\,
\frac{C_A}{2}\frac{1+z^2}{1-z} \nonumber \\
&\times&\frac{2\pi\alpha_s}{N_c}
\overline{I}_{Ap-10,C}(y^-,y_1^-,y_2^-,\ell_T,k_T,x,p,q,z)
 \, , \label{eq:hc10-1} \\
\overline{I}_{Ap-10,C}(y^-,y_1^-,y_2^-,\ell_T,k_T,x,p,q,z)&=&
e^{i(x+x_L)p^+y^-+ix_Dp^+(y_1^--y_2^-)} \,
\theta(-y_2^-)\theta(y^- - y_1^-) \nonumber \\
&\times&e^{-ix_Lp^+y_2^-}
[e^{ix_Dp^+(y^- - y_1^-)/(1-z)}-e^{-ix_Lp^+(y^- - y_1^-)}] \, ,
\label{eq:I10C}  \\
\overline{H}^D_{Ap-10,R}(y^-,y_1^-,y_2^-,k_T,x,p,q,z)&=&
\int d\ell_T^2\frac{ \vec{\ell_T} \cdot
(\vec{\ell_T}-z\vec{k_T}) }
{\ell_T^2 (\vec{\ell_T}-z\vec{k_T})^2}\,
\frac{\alpha_s}{2\pi}\,
\frac{C_A}{2}\frac{1+z^2}{1-z} \nonumber \\
&\times&\frac{2\pi\alpha_s}{N_c}
\overline{I}_{Ap-10,R}(y^-,y_1^-,y_2^-,\ell_T,k_T,x,p,q,z)
 \, , \label{eq:hc10-2} \\
\overline{I}_{Ap-10,R}(y^-,y_1^-,y_2^-,\ell_T,k_T,x,p,q,z)&=&
e^{i(x+x_L)p^+y^-+ix_Dp^+(y_1^--y_2^-)} \,
\theta(-y_2^-)\theta(y_2^- - y_1^-) \nonumber \\
&\times&[e^{-i(x_D-x_D^0)p^+(y_1^- - y_2^-)-ix_Lp^+y_1^-}
\nonumber  \\
& &-e^{-i(1-z/(1-z))x_D p^+(y_1^- - y_2^-) -ix_Lp^+y_2^-}] \, .
\label{eq:I10R}
\end{eqnarray}

In Fig.~\ref{Ap-11} we can make the central cut or the left cut and we
obtain
\begin{eqnarray}
\overline{H}^D_{Ap-11,C}(y^-,y_1^-,y_2^-,k_T,x,p,q,z)&=&
\int d\ell_T^2\frac{ (\vec{\ell_T}-\vec{k_T})\cdot
(\vec{\ell_T}-(1-z)\vec{k_T}) }
{(\vec{\ell_T}-\vec{k_T})^2 (\vec{\ell_T}-(1-z)\vec{k_T})^2}\,
\frac{\alpha_s}{2\pi}\,
\frac{C_A}{2}\frac{1+z^2}{1-z} \nonumber \\
&\times&\frac{2\pi\alpha_s}{N_c}
\overline{I}_{Ap-11,C}(y^-,y_1^-,y_2^-,\ell_T,k_T,x,p,q,z)
 \, , \label{eq:hc11-1} \\
\overline{I}_{Ap-11,C}(y^-,y_1^-,y_2^-,\ell_T,k_T,x,p,q,z)&=&
e^{i(x+x_L)p^+y^-+ix_Dp^+(y_1^--y_2^-)} \,
\theta(-y_2^-)\theta(y^- - y_1^-) \nonumber \\
&\times&e^{-ix_Lp^+(y^- - y_1^-)}
[e^{ix_Dp^+y_2^-/(1-z)}-e^{-ix_Lp^+y_2^-}] \, ,
\label{eq:I11C}   \\
\overline{H}^D_{Ap-11,R}(y^-,y_1^-,y_2^-,k_T,x,p,q,z)&=&
\int d\ell_T^2\frac{ \vec{\ell_T} \cdot
(\vec{\ell_T}-z\vec{k_T}) }
{\ell_T^2 (\vec{\ell_T}-z\vec{k_T})^2}\,
\frac{\alpha_s}{2\pi}\,
\frac{C_A}{2}\frac{1+z^2}{1-z} \nonumber \\
&\times&\frac{2\pi\alpha_s}{N_c}
\overline{I}_{Ap-11,L}(y^-,y_1^-,y_2^-,\ell_T,k_T,x,p,q,z)
 \, , \label{eq:hc11-2} \\
\overline{I}_{Ap-11,L}(y^-,y_1^-,y_2^-,\ell_T,k_T,x,p,q,z)&=&
e^{i(x+x_L)p^+y^-+ix_Dp^+(y_1^--y_2^-)} \,
\theta(y^- - y_1^-)\theta(y_1^- - y_2^-) \nonumber \\
&\times&[e^{-i(x_D-x_D^0)p^+(y_1^- - y_2^-)-ix_Lp^+(y^- - y_2^-)}
\nonumber \\
& & -e^{-i(1-z/(1-z))x_D p^+(y_1^- - y_2^-) -ix_Lp^+(y^- - y_1^-)}] \, .
\label{eq:I11L}
\end{eqnarray}


\end{document}